\documentstyle[12pt,psfig,aaspp4]{article}

\def\hst{{\it HST\/}}

\begin{document}

\title{Toward High-Precision Astrometry with WFPC2. I. Deriving an
Accurate PSF}
\author{Jay Anderson and Ivan R.\ King\\
Astronomy Department, University of California, Berkeley, CA 94720-3411\\
jay@cusp.berkeley.edu, king@glob.berkeley.edu}

\medskip

\begin{abstract}
The first step toward doing high-precision astrometry is the measurement
of individual stars in individual images, a step that is fraught with
dangers when the images are undersampled.  The key to avoiding
systematic positional error in undersampled images is to determine an
extremely accurate point-spread function (PSF).  We apply the
concept of the {\it effective} PSF, and show that in images that consist
of pixels it is the ePSF, rather than the often-used instrumental
PSF, that embodies the information from which accurate star positions
and magnitudes can be derived.  We show how, in a rich star field, one
can use the information from dithered exposures to derive an extremely
accurate effective PSF by iterating between the PSF itself and the star
positions that we measure with it.  We also give a simple but effective
procedure for representing spatial variations of the \hst\ PSF.  With
such attention to the PSF, we find that we are able to measure the
position of a single reasonably bright star in a single image with a
precision of 0.02 pixel (2 mas in WF frames, 1 mas in PC), but with a
systematic accuracy better than 0.002 pixel (0.2 mas in WF, 0.1 mas in
PC), so that multiple observations can reliably be combined to improve
the accuracy by $\surd N$.
\end{abstract}


\clearpage

\section{Introduction}
\label{intro}
Some time ago we began measuring proper motions with the WFPC2 camera of
the {\it Hubble Space Telescope\/} (King et al.\ 1998), and we are now
in the late stages of developing precision techniques for astrometry
with that camera.  Our present level of accuracy has now moved almost an
order of magnitude beyond the level of our previous paper, and we believe 
that it has reached a level that we are unlikely to improve substantially.
This is the first of a series of papers that will describe our
techniques.  

The present paper centers on the vital role of an accurate point-spread
function (PSF), which we find to be the touchstone of accurate astrometry.
Without proper attention to the PSF, astrometry in undersampled images 
tends to suffer from systematic errors which depend on the 
locations of stars with respect to pixel boundaries.  We shall demonstrate 
this pixel-phase error, show that an accurate PSF removes it, and set 
forth methods of determining a PSF of high precision.

Because of the sharpness of its imaging, \hst\ has a unique capability 
for accurate astrometry.  Astrometry of images was initially an 
important selling point of \hst\ (Spitzer 1978, Jefferys 1980, 1985).
It is therefore surprising that ten years after launch there are still 
very few papers in the literature that report significant scientific 
results from astrometry on \hst\ images.  Astrometry has concentrated on
the use of the Fine Guidance Sensors (for one star at a time), but has
made little use of imaging.  There have in fact been several important 
{\it displacement} results for SN1987a, planetary nebulae, stars with 
outflowing jets, and gas clumps in nearby star-formation regions, but 
astrometry proper (i.e., the measurement of stars) has been greatly 
neglected.

It is interesting to ask why it has taken so long to tap \hst's
astrometric potential.  We believe that this delay is due largely to a
natural distrust of astrometry on undersampled images, combined with the
fact that insufficient attention has thus far been devoted to exploring
its possibilities and limitations.  This paper demonstrates that
accurate astrometry {\it can} be carried out on undersampled images, and
that the key is to have an extremely accurate PSF.

Differential astrometry can be divided into two distinct yet not
entirely independent tasks:
(1) measuring positions of individual stars in individual images and
(2) comparing these positions with positions measured for the same 
    stars in other images.
In this paper we will discuss the first of these tasks.  The second,
which involves finding accurate coordinate transformations from
one frame to another, we will take up in a subsequent paper.  We will
focus on measurements made with \hst's WFPC2 camera in rich fields
such as one finds in globular clusters, but many of our precepts and
methods are applicable in broader contexts.

We begin with a brief discussion of the impact undersampling has on
astrometry.  In Section \ref{undersam} we highlight the most significant
impediment to accurate astrometry:\ pixel-phase error, a systematic
error that is related to the location of a star with respect to pixel 
boundaries.  We trace the source of this systematic error to inadequate 
modeling of the PSF.  We then (Sect.\ \ref{nature}) show how treating the 
{\it effective\/}, rather than the instrumental, PSF, allows us to 
arrive at an accurate model which can be used to measure unbiased 
positions for stars.  In Section \ref{find_psf} we outline our
procedure for obtaining such accurate PSFs and using them to measure
star images.  We also demonstrate the accuracy of our measurements
and examine the time stability of the PSF.  In Section~\ref{dith} 
we make recommendations for how best to dither future observations 
in order to facilitate an accurate PSF determination.  Finally, in 
Section \ref{ePSFextend} we show how our effective-PSF treatment can be 
extended to other contexts.

\clearpage

\section{Undersampled images and astrometry} 
\label{undersam}
The point-spread function formed by \hst\ has a core that is about
50 milli-arcsec (mas) wide, the exact size depending on wavelength, of
course.  Since the pixel sizes of the CCDs in the Planetary Camera (PC)
and the Wide Field Camera (WF) are 45 and 100 mas respectively, it is
clear that even the PC is undersampled, and the WF is badly so.

Adequate sampling is conventionally thought of in connection with issues
of resolving power, such as the separation of two close point sources or
the discernment of fine detail in extended objects.  In \hst\ images of 
globular clusters, however, essentially all of the objects are relatively 
isolated point-source stars.  Unlike an extended object, a point source 
needs only three parameters to describe it fully:\ its center $(x,y)$ and 
its total flux ($f$).  The whole task of photometry and astrometry, then, is 
to derive this triplet of parameters from the array of pixels that constitute 
the image of a star.

The effects of undersampling upon this task are not as serious as one
might initially suppose.  For instance, we do not require Nyquist
sampling in order to measure precise positions for point sources.  
We simply require that the PSF not be so sharp that {\it all} of a star's
flux falls within a single pixel.  As long as a reasonable amount of
flux falls in the surrounding pixels, so that shifting the star's
position by a small amount would redistribute light from one pixel to
another, we have the basic information that we need to measure accurate
positions.  In the WF, when a star is centered on a pixel, that pixel
receives about 40\% of the star's light, while the four directly adjacent 
pixels each get about 8\%.  This spill-over of the PSF core into the 
adjacent pixels is sufficient to enable us to pinpoint the star's 
location within the brightest pixel. 

Such measurements are not without complications, however.  While it is
theoretically possible to obtain accurate positions in undersampled
images, such measurements demand an accurate model of the PSF.  Any
inaccuracies in the PSF model will induce systematic biases in the
measured positions.  To illustrate this basic point, in
Figure~\ref{simple_ex} we fit the same 1-D stellar profile with two
different models of the PSF.  The stellar profile is represented by the
histogram, which gives the values of the closest three pixels to the
center of this star.  The two PSF models (both of which integrate to fit
the pixel values exactly) yield estimates for the ``center'' that are
strikingly different, compared with our accidental error of measurement,
which is about 0.02 pixel.  It is obvious that the bias induced depends 
on the location of the star within the pixel.  We refer to this bias as 
{\it pixel-phase error}.

[FIG.\ 1 HERE]

In conjunction with its implications for centering, Fig.~\ref{simple_ex}
also demonstrates that since the same undersampled stellar profile can
be fit by drastically different PSFs, it will not be easy to determine
which one of the many consistent PSFs is actually the ``right'' one for 
the image in question.  If we somehow knew the position of the star, this 
might tie down the PSF model; however, we clearly cannot know the position 
of the star accurately without recourse to an accurate PSF.  

The breaking of this degeneracy is the major challenge of undersampled
astrometry.  In short, we will tease apart the shape of the PSF and the
positions of stars by incorporating observations made of the same star 
at different dither positions (different placements with respect to pixel 
boundaries).  This iterative process is the major contribution of this paper.

We note also, in passing, that an alternative approach might be to 
try to remedy the undersampling, following, for example, the Fourier
image-reconstruction methods of Lauer (1999a).  We prefer instead to 
aim directly at deriving a correct PSF, which allows us to work with 
the original data without modifying the pixels in any way; this 
simplifies the task of understanding the errors, which is extremely 
important to our objectives.

\clearpage

\subsection{The present data set}
\label{data_set}
It was in our early attempts at astrometry that we were led to see the
crucial role of an accurate PSF in the measurement of star images.
These efforts concentrated on a set of 15 images of the center of the
globular cluster 47 Tucanae taken through the F300W filter, provided to us by 
Georges Meylan, the P.I.\ of \hst\ program GO-5912.  The images, rich 
but not crowded, have identical exposure times, have nearly identical 
PSFs, and are well dithered.  

In preparing these images for analysis, we adjusted each pixel value
according to the true area of the pixel, in order to remedy the errors
that are introduced by the flatfield correction made by the standard
pipeline.  Specifically, the pixel-area correction is for optical
distortion (Biretta et al.\ 1996) and for the 34th-row anomaly (Anderson
\& King 1999).  (For the prescription, see Biretta et al., Sec.\ 5.10;
for an explanation of why the correction is needed, see the fifth
paragraph of Anderson \& King.)  We made no attempt to remove cosmic
rays from the individual images.  The intercomparison of the many
individual photometric and astrometric measurements that we have for
each star allows us easily to discard, at a later stage of the analysis,
those measurements that are infected by cosmic rays.

The WFPC2 detectors on board HST have become increasingly subject to
charge-transfer-efficiency (CTE) losses (Whitmore, Heyer, \& Casertano
1999 and Riess, Biretta, \& Casertano 1999).  While not all the
manifestations of CTE are fully understood, the major symptom is that
when the background is low, stars at the top of a chip (and, to a lesser
extent, those at the right) appear fainter than those at the bottom (and
left).  The standard remedy is to apply a simple ramp correction to the
measured fluxes of stars, proportional to $(1.0 + C_y \times y/800 + C_x
\times x/800)$, where $C_y$ is typically between 2\% and 8\%, and $C_x$
is typically half that.  In principle this correction should be applied
to each pixel of an image before analysis, but in practice the
differential correction across the image of a star is negligible, so
that we can apply the CTE corrections at the final stage of our
photometry, while our astrometry is unaffected by it.

CTE losses are probably more complicated than the simple ramp model
allows.  Losses could easily be different for each star depending on its
own brightness and the specific distribution of stars downstream from
it.  We have examined our astrometric data for such ``shadowing''
signatures, but failed to find any significant trends.  We will present
the details of these tests in a subsequent paper.

\subsection{Recognizing pixel-phase errors} 
\label{rec_ppe}
With the flux-normalized images in hand, we attempted to make the
most accurate positional measurements that we could on the individual
stars in each image, by deriving a PSF from the image itself and fitting
it by least squares to each star image.  We profited very much from
Meylan's foresight in taking a large set of well-dithered images.  With
multiple observations of the same star, we could find a mean position
$(\bar{x},\bar{y})$ for each, and examine the 15 residuals
$(x_n\!-\!\bar{x},y_n\!-\!\bar{y})$.  These residuals indicated that we
were measuring positions with an accidental error of about 0.02 pixel.
More important, these residuals also gave us a chance to examine our
measurements for systematic errors.

(Note:\ calculating the mean position for each star requires transforming 
measured positions from the frame of each image into a common frame, a process 
that introduces many subtle problems that will be discussed in our later
papers.  For our present purpose, however, it suffices to know that the
mean position of a star can indeed be calculated.)

Knowing that inaccuracies in our adopted PSF model might lead to
pixel-phase error, we examined the residuals as a function of where each
stellar image was located with respect to pixel boundaries---what we
call its pixel phase, expressed quite simply by the fractional part of
its position measurement: $\phi_x \equiv x - {\rm int}(x+0.5)$.  Figure
\ref{ppe} shows the result of this examination.  The nine panels show
stars in 9 sections of the image, which we examined separately because
the PSF is known to vary spatially in WFPC2 images.  (In the present
illustration, the same PSF was used in all of the 9 regions, but it was
determined only from stars in the central region.)

[FIG.\ 2 HERE]

The trends in the figure are so clear because the figure contains the
residuals from 3000 well-exposed stars in 15 images---45,000 residuals
in all.  We were shocked by the size of the trends, which are comparable
to or larger than the accidental errors.  The trends were similar in
size for the $y$ residuals against $\phi_y$, but the cross trends ($x$
residuals vs.\ $\phi_y$ and $y$ residuals vs.\ $\phi_x$) were
considerably smaller.  The trends for the PC chip were similar in nature
to those for the WF chips.  

(The apparent independence of errors in $x$ and in $y$ is probably
illusory.  It is likely that errors would be approximately symmetrical
around the center of the pixel, and such a symmetry would produce that
illusion.)

Given these clear trends, a possible course would have been to use these
measurements to calibrate out the systematic error empirically; but we
chose instead to look for the source of the error, with the aim of
eliminating it.  Paradoxically, the severe undersampling of WFPC2 makes
our task easier, because it makes the pixel-phase error so large that
its cause is more readily evident.

From the considerations in the introduction to the present section, we
know that even though stellar images appear to be well fit by a given
PSF, the positions measured may nevertheless be systematically biased.
Since we did not measure accurate positions even in the central region
of the image, where the PSF was determined, it is natural to suspect the
PSF model as the source of this error.

Before setting out to find a better PSF, however, we first investigated
other easier-to-fix possibilities.  One step was to verify that given an
accurate PSF, our routines could indeed measure good positions.  We did
this by inserting artificial stars made with our model PSF, at random
locations in the image, and adding a suitable amount of noise.  We
measured them with the (correctly known) PSF, and found that we
recovered them without pixel-phase error.

We then arbitrarily adjusted our PSF model and found that by making its
core narrower or broader we could change the amplitude of the
pixel-phase error in measurements of the actual data.  It was becoming
clear that the problem lay with our PSFs.  In addition, in the hope of a
better PSF-finding algorithm, we reduced the data with DAOPHOT, but
found almost identical pixel-phase errors.

It makes sense that astrometry should make very different demands on a
PSF from those of photometry.  We can make a good measurement of the
total flux of a relatively isolated star by simply adding up the flux
within a chosen aperture and correcting that sum according to what
fraction of the PSF should fall within that aperture.  It matters little
if the PSF mis-predicts the exact pixel-by-pixel distribution; the only
necessity is that the {\it sum} of the PSF model over the aperture be
equal to that of the true PSF.  Astrometry, on the other hand, depends
entirely on the {\it differences} in adjacent pixel values; this means
that the ability of the model to predict each individual pixel value is
of paramount importance to measuring star positions.  Slight errors in
the PSF will impact astrometry directly, but photometry only indirectly.

While these considerations clearly demonstrate the need for an accurate
PSF, they do not tell us {\it how} to obtain one.  In our quest for
a better PSF, we found that it is worth while not only to reconsider
how we measure the PSF, but also to ask the more basic question of 
{\it what} PSF we should be seeking in the first place.

\clearpage

\section{Reconsidering the PSF}
\label{nature}
It is clear that if we desire to model the PSF as accurately as
possible, we must take full account of the pixellated nature of our
image data.  In this section we will examine the impact of pixellation
on the PSF problem, and will show that there is a very straighforward
way of coping with it empirically, which leads to a representation of
the PSF that has great practical advantages.  We will then discuss the
nature and the advantages of this empirical representation, for which we
reserve the name ``effective PSF," used with a very specific meaning.
Finally, we will contrast our treatment with other approaches to PSF
modeling.  We wish to emphasize again that the motivation for this new
approach is the need to model the images of stars more accurately than
has heretofore been possible.

\subsection{The role of pixels in the PSF}
\label{pixelPSF}
The PSF---the two-dimensional spread of light that an instrument 
produces when its input is a point source---is central to the art and 
science of image processing.  Each star in an astronomical image is a
replica of the PSF, distinguished only by its position and its flux.
The intuitive nature of the PSF is obvious:\ it is the profile that the
instrument renders when the input is a point source.  In practice,
however, the nature of the PSF is more subtle.  The telescope produces,
at the focal plane, what we may call an instrumental PSF (iPSF).  But
one never directly observes this PSF; what one sees instead is the array
of pixel values that results from it.  Even with an infinite signal-to-noise
ratio and a perfect flat field, the same star will produce
different arrays of numbers depending on where its center falls with
respect to pixel boundaries.

The astronomical image produced by a detector is indeed only a set of
numbers, each of which is the number of counts registered by an
individual pixel.  If we examine the source of one of these pixel
values, it is clear that it results from integrating the instrumental
PSF over the two-dimensional sensitivity profile of a pixel.  
%%
% Following sentence (old version commented out) has been replaced by
% the version that follows.
%
%The latter
%resembles a square top-hat function (sometimes referred to as a $\Pi$
%function), but it need not be exactly a $\Pi$ function, and usually is
%not.
%  
One might
expect the latter to resemble a square top-hat function, but in fact it
does not (Lauer 1999b).  
Not only does sensitivity vary within the pixel; there is also a 
tendency for photoelectrons to diffuse out of the pixel (Holtzman
et al.\ 1995, Krist \& Hook 1999, Lauer 1999b).  The Lauer paper gives
examples of intra-pixel sensitivity maps (the function that we call
${\cal R}$ below).  We note that it is characteristic of many detectors 
that the sensitivity profile differs only inconsequentially from pixel 
to pixel (see, for example, Shaklan, Sharman, \& Pravdo 1995); we 
confine our discussion here to detectors for which this is the case.

\subsection{The effective PSF}
\label{eff_psf}
Thus every pixel value in a star image is the result of an integration
over a pixel centered at some offset $(\Delta x,\Delta y)$ from the center 
of the iPSF.  Everything that we ever observe directly consists of such 
pixel values; we never directly encounter the iPSF itself.  
As our initial impression was that we need to model {\it that} PSF as
accurately as possible, it was frustrating to realize that we can never 
observe it directly---until we recognized that we do not {\it need} to 
do so.  

The hidden nature of the iPSF becomes clear when we express mathematically
what we {\it do} observe. If in the vicinity of pixel 
$(i,j)$, which is centered by definition at $x=i$, $y=j$, there is a
point source centered at $(x_*,y_*)$, then the 
flux in that pixel will be 
\begin{equation}
P_{ij} = f_* \int_{-\infty}^{\infty} 
             \int_{-\infty}^{\infty}
                  {\cal R} (x\!-\!i,y\!-\!j) 
                  \psi_I(x\!-\!x_*,y\!-\!y_*) \, dx\, dy + s_*,
\label{ipsf2pix}
\end{equation}
where $f_*$ is a flux factor that expresses the brightness of the star,
$\cal R$ is the two-dimensional sensitivity profile of a pixel,
$\psi_I(\Delta x,\Delta y)$ is the instrumental PSF, specifically, the
fraction of light (per unit pixel area) that falls on the detector at a
point offset by $(\Delta x,\Delta y)$ from the star's center
$(x_*,y_*$), and $s_*$ is the value of the background at that pixel.

We shall show that this pixel value is the result of a convolution
between the iPSF and the sensitivity profile of a pixel.  To do this we
note first that because the integration in Eq.\ (\ref{ipsf2pix}) is over
an infinite range, we can make a shift of zero point and write it as
\begin{equation}
P_{ij} = f_* \int_{-\infty}^{\infty} 
             \int_{-\infty}^{\infty}
                  {\cal R} (x,y)
                  \psi_I(x+\Delta x, y+ \Delta y) \, dx\, dy + s_*,
\label{ipsf2pixsh}
\end{equation}
where we have written $\Delta x$ and $\Delta y$ in place of $i-x_*$ and
$j-y_*$.  Noting that the variables of integration $x$ and $y$ are dummy
variables, we replace them by $-x$ and $-y$; the equation then becomes
\begin{equation}
P_{ij} = f_* \int_{-\infty}^{\infty} 
             \int_{-\infty}^{\infty}
                  {\cal R} (-x,-y)
                  \psi_I(\Delta x-x, \Delta y-y) \, dx\, dy + s_*.
\label{ipsf2pixshr}
\end{equation}

We now define the convolution
\begin{equation}
\psi_E(\Delta x, \Delta y)  \equiv
     \int_{-\infty}^{\infty} 
     \int_{-\infty}^{\infty}
      {\cal R}^\prime (x,y)  \psi_I(\Delta x-x,\Delta y-y)\, dx\, dy,
\label{epsfconv}
\end{equation}
where ${\cal R}^\prime (x,y) = {\cal R} (-x,-y)$.
It is easy then to see that Eq.\ (\ref{ipsf2pixshr}) is equivalent to
\begin{equation}
P_{ij} = f_* \psi_E(i\!-\!x_*,j\!-\!y_*) + s_*. 
\label{epsf2pix}
\end{equation}
We call $\psi_E(\Delta x,\Delta y)$ the {\it effective PSF} (ePSF),
because it gives directly, through Eq.\ (\ref{epsf2pix}), the fraction
of a star's light that should fall in each pixel of a star image,
according to where the center of that pixel lies relative to the center
of the star, i.e., the offset $(\Delta x, \Delta y)$.  

Note that $\psi_E$ is a continuous function and is
smoother than the hidden $\psi_I$.  It is worth restating that, since 
the only data we have are the pixel values recorded by the detector, 
$\psi_E$ contains all the information we will ever need to know about 
$\psi_I$ and $\cal R$.  In other words, Eq.~(\ref{epsf2pix}) says that 
we never need to know $\psi_I$ and $\cal R$ separately.

The reader should note that whereas other writers use the term
``effective PSF'' in a general way to refer to a PSF that has been
integrated over pixels, in this paper the term always has the explicit
meaning that has just been defined and described.

\subsection{Ways to think about the effective PSF} 
\label{think_epsf}

Since the effective PSF plays such an irreducible role in image analysis, 
it is valuable to develop an intuitive understanding of what it can 
tell us about how point sources appear in an image.  Figure~\ref{ins2eff} 
shows a graphical example of an undersampled instrumental PSF 
(short-axis FWHM $\sim$ 0.8 pixel) and the corresponding effective PSF.  
At any point (the cross), the ePSF is equal to the integration of a
pixel (solid box) centered at that point over the instrumental PSF.
Both functions are quite smooth, but the ePSF is, naturally, less 
concentrated than the iPSF.  

[FIG.\ 3 HERE]

From Eq.~(\ref{epsf2pix}) we see that the central point of the ePSF, 
$\psi_E(0,0)$, point ``C'' in the figure, tells us directly what 
fraction of a star's light falls within the star's central pixel when the 
PSF is centered on that pixel.  Similarly, $\psi_E(1,0)$, the value of
the ePSF at point ``R'', tells us what fraction of a star's light falls 
within a pixel whose center is exactly one pixel to the right of the star's
center.  In general, if the center of a star is at $(\delta x,\delta y)$ 
relative to the center of a pixel, then $\psi_E(0\!-\!\delta x,0\!-\!\delta y)$
gives the fraction of light that should fall in the star's central pixel, 
and $\psi_E(1\!-\!\delta x,0\!-\!\delta y)$ gives the fraction received in 
the first pixel to the right of the center.  The gradient of $\psi_E$ tells
us the rate at which flux is transferred into or out of a pixel as 
$(\delta x, \delta y)$ are varied.

[FIG.\ 4 HERE]

Figure~\ref{coords} shows the relation between the pixels in a stellar 
image and the points in the effective PSF that they correspond to.  We show 
the center of the star as the open circle.  Each pixel of the stellar image 
samples the ePSF at a different location; the array of pixels that make 
up a star's image sample the ePSF at the array of points marked by the crosses.

\subsection{Advantages of the effective approach}
\label{advant}
There are three major advantages to treating the effective PSF directly, 
rather than working via the instrumental PSF, when analyzing images.

The first advantage is simplicity.  In fitting the ePSF to a stellar image,
no integration is needed.  We simply {\it evaluate} the ePSF at the position 
of each pixel of the star image (relative to the presumed center) and scale 
it by the flux factor appropriate for the star.  The fitting process,
correspondingly, consists of adjusting the values of $x_*$, $y_*$, and $f_*$ 
until the sum of the squares of the residuals,
\begin{equation}
  \sum_{ij} w_{ij} \biggl[ (P_{ij} - s_*) - f_*
  \psi_E(i\!-\!x_*,j\!-\!y_*) \biggr]^2 \> ,
\label{fitpix}
\end{equation}
is minimized.  Again, this procedure involves no integrations over
pixels, only evaluations at discrete points.  The fitting is simply a
least-squares solution for the optimal parameters in the fit of a
continuous function to a given set of points.  Section~\ref{fitPSF}
describes this procedure in detail.

The second advantage is that it is much easier to solve for an effective
PSF than for an instrumental PSF.  Every pixel of every star image samples 
the effective PSF at one discrete point.  If we turn Eq.~(\ref{epsf2pix}) 
around, we see that the ePSF can be estimated by
\begin{equation}  
    \hat{\psi}_E(\Delta x, \Delta y) = { P_{ij} - s_* \over { f_* } }.
\label{epsfsam}
\end{equation}  
So if we know $x_*$, $y_*$, and $f_*$, they then tell us at which offset,
$\Delta x = i\!-\!x_*$ and $\Delta y = j\!-\!y_*$, the pixel at $(i,j)$ has 
sampled the effective PSF and how this sampling has been scaled.  If we 
know the set $\{x_*,y_*,f_*\}$ for each of $N$ stars, and take 5$\times$5 
pixels about each of them, then we have 25$\times N$ samplings of the 
effective PSF, each at a point in the ePSF.  With 3000 stars in each 
of 15 images, we get well over a million estimates of 
$\psi_E(\Delta x,\Delta y)$ at various offsets in this 5$\times$5-pixel 
area. Our procedure for representing and deriving a smooth ePSF from this 
wealth of data will be described in detail in Section~\ref{epsf_const}.

When the PSF is undersampled, it is especially useful to be able
to work with such a directly observed function.  By contrast, the 
conventional practice of trying to deduce the hidden instrumental PSF
(or the major part of it, as in the underlying Gaussian that DAOPHOT
uses) from observed pixel values (essentially a deconvolution), and then 
re-integrating over $\cal R$ to fit it to the same pixels, appears to 
be a roundabout and counter-productive procedure.  As errors can arise 
in both of these steps, it is clear that the effective-PSF approach is 
not only easier, but numerically preferable.

The third advantage of the ePSF is more subtle, and in many ways more
valuable.  Its values result from integration over the actual pixel
sensitivity profile of the detector, whatever it may be.  Since, as we 
have noted above, we have no need to untangle the contributions of 
$\psi_I$ and $\cal R$, we do not need to make any assumptions about 
how sensitivity varies within a pixel.  Our effective PSF simply (and 
accurately) represents whatever results from the combination of the 
detector and instrumental PSF, with no assumptions on our part.

\subsection{Relation to other treatments}
\label{relation}
The concept and use of the effective PSF that we have introduced here
are quite new when considered as a whole, but they are related to, and
in some ways build on, other treatments that have been made.  The
salient characteristics of our ePSF are that (1) it is completely
empirical, without reliance on any analytic approximation, (2) it is
derived directly from observed pixel values, without any modification
other than scaling, and (3) fits to pixel values follow directly from 
it by simple evaluation and scaling, without any integration.

We were originally stimulated to the concept of the effective PSF by
ideas contained in Lauer's discussion of how to use Fourier techniques 
to combine multiple undersampled dithered images into a single well-sampled
``superimage'' (Lauer 1999a).  Although the concept that we have 
introduced here is not explicitly stated in Lauer's paper, we believe 
that it is implicit in his discussion.  But we have moved in a different 
direction, because our aims are different.  Lauer's desire is 
to create a finer-scale ``superimage'' that is equivalent to the one
that would have been produced by a detector that is well sampled, and
his method of doing this is direct, elegant, and probably the best 
treatment available for extended objects.  Such a well-sampled 
superimage is well suited for deconvolution, which will produce an 
estimate of the true scene.  Deconvolution is not, however, a tool 
that is relevant to the measurement of point sources.

While it is indeed possible to measure positions of stars on such a
well-sampled superimage, to create such an image in the presence of
geometric distortion requires an intimate knowledge of the inter-image
transformations, which can be attained only by accurately measuring
stars on the images independently.  Furthermore, one of the major
objectives of our project will be the accurate assessment of our errors,
which is straightforward only if we have multiple independent
measurements of each star's position, which we obtain by measuring each
image separately and combining the positions afterwards.  It is the
intercomparison of such measurements that allows us to examine our data
for systematic errors such as pixel-phase error or uncorrected
distortion.  We therefore find it far preferable to develop an accurate
model of the effective PSF, as we defined it in Section \ref{eff_psf},
and to use it to measure a position for each star in each image.

We are by no means the first to create a PSF that has been integrated
over pixels.  It is this very idea that is so interwoven into the Lauer
paper to which we have just referred.  And integration over pixels is
included, in one way or another, in DAOPHOT (Stetson 1987), DoPHOT
(Schechter, Mateo, \& Saha 1993), and most other procedures that fit
star images with a PSF.

It is DAOPHOT that comes closest to the idea of our effective PSF.
After fitting the closest Gaussian\footnote{In recent versions, DAOPHOT
has allowed the substitution of a Moffat function for the Gaussian, but
the principle is the same.}, DAOPHOT creates its PSF by integrating the
Gaussian, correctly placed, over the area of each pixel in a star, and
taking the difference between this result and the observed value of that
pixel.  It tabulates these residuals, evaluated at half-pixel
intervals starting at the center of the Gaussian.  To fit the PSF to a
star, for each pixel it integrates the Gaussian and then does a cubic
interpolation in its array of residuals to find the correction that
needs to be added to it.  Thus DAOPHOT operates in a hybrid manner.  It
does treat its array of residuals as an adequately fine tabulation of a
continuous function that is already integrated over pixels, but it still
needs to integrate the Gaussian over each pixel.

DAOPHOT, as we have indicated, suffers from astrometric pixel-phase
error.  The reason is that it lacks our procedure of correcting
positions by dithered averaging and iterating between PSF shape and star
positions, which we will describe in Section \ref{find_psf}.
It is not surprising that DAOPHOT does not produce the best possible
positions; it was designed to do good photometry, not astrometry.  To
paraphrase a remark made to us by its author, Peter Stetson, a tool used
for the wrong purpose is not likely to be optimal.

DoPHOT (Schechter et al.\ 1993) presents an interesting contrast.  It
ignores the pixel-integration problem and simply evaluates its PSF at
the center of each pixel that is to be fitted (Saha, private
communication).  At first glance this might seem to imply the use of an
iPSF, without integration over the pixel.  In fact, however, DoPHOT
derives its PSF from pixel values, so that its PSF has indeed been
integrated over pixels, by the detector itself.  Schechter et al.\ show
that DoPHOT produces quite good photometry; we surmise that the good
performance in spite of such a simplified approach to pixels is due to
the fact that its PSF is derived from pixel values in exactly the same
way in which it will be used on them.  We have no doubt, however, that
the pixel-phase errors in the positions that it derives are quite serious.  
Unlike DAOPHOT, DoPHOT does not seem amenable to the average-and-iterate 
procedure; its purely analytic PSF representation lacks the necessary 
flexibility.

We adopt here a purely empirical model, akin to using a DAOPHOT-type 
lookup-table without the benefit of an analytical model.  Stetson (1987) 
considers such a model, but adopts the analytical backbone in order to 
minimize the undersampling and interpolation concerns.  Our finer sampling
combined with our PSF-construction procedure will automatically address
these concerns.  

Such a tabular approach has the obvious advantages of flexibility and
simplicity.  Because we observe the effective PSF directly in each star 
image (see Eq.~\ref{epsfsam}), and can see exactly which part of the PSF
needs adjusting to better fit the observations, we can directly adjust 
the grid to better reflect the observed stellar profiles.  Such a 
grid-representation also is particularly amenable to the modeling of 
``ugly'' PSFs (such as intentionally or unintentionally trailed images) 
which might not have an obvious analytical backbone.  As our goals involve 
modeling the PSF to unprecedented accuracy, the flexibility of the tabular 
approach is a very important factor.

\subsection{Details of our ePSF model}
\label{ePSFdetails}
The effective PSF is a continuous function which extends out to several 
hundreds of WF pixels.  In the undersampled WFPC2 cameras, practically all 
of the information from which we can fix the position and flux of a star 
is contained in its central 5$\times$5 pixels, so we will seek a numerical 
representation of that part of the PSF.  

As we mentioned above, we will represent the ePSF by specifying its value 
at a grid of points.  After some experimentation we found that it was 
adequate to tabulate it with four grid points per pixel width.  (This 
choice is purely a matter of convenience, to represent this continuous 
function well enough that we can easily interpolate values at intermediate 
points.  Since the ePSF does not have pixels, it should be clear that the 
grid spacing has nothing to do with any attempt to subsample an image.)  
Our aim is thus to tabulate values of the ePSF at an array of 21$\times$21 
points (21 rather than 20 because we place points on the boundaries).  
When we need to evaluate the ePSF at locations between the grid points 
(for example, when fitting stars), we will use bi-cubic spline interpolation.

The extreme flexibility of such a grid-based model requires us to adopt
some conventions regarding how we will center and normalize the PSF.  

The definition of the center of any PSF is of course arbitrary.  If the
PSF is asymmetrical, one may ask if one should choose the peak, the
centroid, or what?  Rather than use either of these more conventional
approaches, we decided to center our grid in a way that flows naturally
from the pixel-integrated nature of the ePSF.

Imagine taking a star that is centered on a pixel and moving it to the
right, across the detector.  The value of the ``central'' pixel goes
down as that of the one to the right goes up.  At the point where the
two pixels are exactly equal we declare that the center is on the
boundary between the two pixels, so that (according to
Fig.~\ref{coords}), the central pixel is sampling $\psi_E(-0.5,0.0)$ and
the one to the right is sampling $\psi_E(0.5,0.0)$.  We adopt a similar
definition for the $y$ center.

A major benefit of this convention is that it is easy to re-center a
PSF, since the gradient is nearly maximal at 0.5 pixels from the center
and the condition for centering is well defined, namely that 
$\psi_E(-0.5,0.0) \equiv \psi_E(0.5,0.0)$.  A peak-based centering 
algorithm is not so well constrained, since the gradient is zero at 
the center and the determination of the peak depends sensitively on the 
innermost grid points.  We show in Section~\ref{sam2psf} how our 
centering convention is enforced.  

A useful consequence of this convention is that, noise considerations aside, 
the brightest pixel of a star will be drawn from the inner pixel-sized 
region of the PSF, namely $\Delta x$ between $-0.5$ and 0.5 and $\Delta y$ 
between $-0.5$ and 0.5.  We can now further identify (for example) the
region between a $\Delta x$ of $-1.5$ and $-0.5$ and a $\Delta y$ of $-0.5$
and $0.5$ to be the region of the ePSF corresponding to the next pixel
to the left of the brightest pixel.  The behavior of the ePSF at different 
points in this region reflects how the fraction of light falling in this 
left-of-center pixel varies with the pixel phase of the star.  The other 
pixel-sized regions of the ePSF (designated by the dotted lines in 
Fig.~\ref{ins2eff} and in the right panel of Fig.~\ref{coords}) have 
similar interpretations.  It is easy to see that the ePSF will be perfectly 
continuous across all ``pixel boundaries,'' as Eq.~(\ref{epsf2pix}) implies 
and Fig.~\ref{ins2eff} demonstrates.

It is standard to normalize PSFs to have unit volume,  so that the flux 
factor in Eq.~(\ref{epsf2pix}) will indeed correspond to the total flux of 
the star.  Lauer (1999b) shows that in undersampled detectors (such as WFPC2 
or NICMOS3), the integrated flux recorded by a star can vary significantly 
as a function of where the star's core falls in the central pixel.  In WFPC2, 
the top and bottom edges of pixels are about 1.5\% more sensitive and the 
left and right edges about 1.5\% less sensitive than average (this depends 
somewhat on the filter).  We therefore normalize our ePSFs so that a star 
of unit flux that is centered on a pixel will have a volume of unity in 
its central 5$\times$5 pixels (the region over which we will be tabulating 
the ePSF, below); a unit-flux star that is centered differently will 
then have a volume that is not unity but that correctly expresses the 
differences in the apparent brightness of a star that result both from 
the non-uniformities in the pixel-response function and from the differing 
fraction of total light that falls within the 5$\times$5-pixel aperture.  
That this is so is a natural consequence of the way in which we have 
defined the ePSF; notice again that we have taken the properties of the 
pixel-response function into account without needing to know that function 
explicitly.

In summary, the ePSF is a continuous function which tells us what
fraction of a star's light should fall within any pixel, according to
where that pixel lies with respect to the star's center.  We will represent
this ePSF by tabulating its value at an array of points that is more finely
spaced than the pixels themselves.  We will use bi-cubic interpolation to
evaluate the ePSF at intermediate locations.  We will center our grid in
such a way that the brightest pixel of a star will be drawn from the 
central-pixel region of the ePSF.  Finally, our representation is naturally 
able to account for variations of sensitivity within a pixel.

\clearpage

\section{Deriving an accurate PSF}
\label{find_psf}
The effective-PSF approach illustrated above clearly leads to a more
direct way to analyze images.  Nonetheless, modeling a PSF
``effectively'' will not automatically produce a more accurate PSF.  The
fundamental problem of undersampled star images remains: namely, the
degeneracy between the shape of the PSF and the position that it finds
for each star image.

We will describe below our procedure for determining an accurate
effective PSF for the WF2 chip in Meylan's first-epoch set of images.
As we indicated earlier, we will be tabulating values of the ePSF at an
array of 21$\times$21 points.  It should be noted that our discussion 
will refer, of course, to the derivation of the PSF for a single chip; 
each of the four chips must be treated separately.  A final point to note 
is that we are tacitly assuming here that all of the exposures that we 
are combining have the same PSF.  This assumption will be justified 
empirically in Section \ref{indiv_comp}.

\subsection{The need for iteration} 
\label{epsf_iter}
Whether the image is undersampled or not, the construction of an 
accurate PSF is necessarily an iterative procedure:  we cannot 
derive a PSF from an image without prior knowledge of star positions 
and fluxes, nor can we measure positions and fluxes without recourse 
to a PSF.  In practice we  will alternate between deriving the PSF and 
measuring the stars, improving both with each iteration.  Undersampling 
makes iteration all the more crucial, since we must in addition use multiple 
ditherings to tease apart the degenerate PSF-shape and star-positioning 
information.

Before beginning the iterative process of solving for a PSF, we must
first decide which stars can tell us something about its shape.  We
choose all those isolated stars that have a $S/N$ greater than $\sim$10
in each of their inner 5$\times$5 pixels.  This amounts to all isolated
stars (i.e., those having no brighter neighbors within a radius of 5 or
so pixels) that have more than 250 DN ($\sim$1750 detected photons) in
the 5$\times$5-pixel area.  In each of our images roughly 3000 stars
qualify.

(We should mention, by the way, that crowding is never a problem for us.
Since the astrometric results toward which we are moving never require
completeness, accuracy being our only aim, we can throw out any stars
that suffer from crowding.  This is easily done, even when the
interfering neighbor is not directly detected; we have found that a star
that suffers from crowding can invariably be recognized by the fact that
its measurements have larger residuals than those of a star of
comparable magnitude that is uncrowded.)

For each of these stars, we require an initial estimate of its flux and 
position and background.  For this we could use DAOPHOT with a library PSF, 
or even aperture photometry, for fluxes, and simple centroids for positions.  
All that is really needed is a reasonably good set of starting values.
We adopt a sky value for each star by taking the mode of the pixels in an 
annulus with an inner radius of 4 pixels and an outer radius of 7 pixels.

Each PSF-finding iteration consists of three major stages:  
(1) We convert each pixel value in the image of each star into an 
    estimate of the corresponding point in the effective PSF, using 
    the most recent values of the star's flux and position in
    Eq.\ (\ref{epsfsam}).  In the present data set 
    this corresponds to more than a million samplings of various points 
    of the effective PSF.  We then construct a single effective PSF from 
    this multitude of point-samplings.  (Note that in practice  we will 
    end up dividing the image into 9 separately treated sections 
    in order to represent the spatial variation of the PSF, 
    but even so each of the 9 PSFs will be based on well over $10^5$ 
    samplings.)
(2) We use this effective PSF to remeasure, by least-squares fitting, a 
    position and a flux for each star in each image.  
(3) We average together the multiple observations we have for each 
    star from the 15 different pointings, to obtain positions and 
    fluxes that are more accurate (in a random sense and now, as a
    result of combining the various dithers, in a systematic sense too) 
    than the individual measurements.  We then transform each of these average
    positions back into the frame of each individual image so as
    to place each individual-pixel sampling more accurately in the ePSF
    in step (1).  It is, in fact, this averaging step that is the key to 
    breaking the degeneracy that leads to pixel-phase error.  

The three stages are explained in detail in the following subsections.
The overall procedure typically converges in four or five iterations,
although the need to converge 9 separate PSFs at once has led us to
continue as far as a 12th iteration.

\subsection{Stage 1:  The construction of the effective PSF} 
\label{epsf_const}
We pointed out in Section~\ref{eff_psf} that every pixel in an image of a star 
samples the effective PSF at {\it one} point, provided we know
what the star's position and flux are.  Eq.\ (\ref{epsfsam}) gives the
relationship mathematically, and Fig.~\ref{coords} shows the geometrical
relationship for the pixels of one star.  What makes the derivation of
an accurate ePSF possible is that every reasonably bright star
contributes one sampling per pixel per image.

\subsubsection{From many samplings to one PSF}
\label{sam2psf}

Figure~\ref{datapts} simulates the samplings from 500 stars in the ePSF,
along with the grid points at which we evaluate the ePSF.  (Note that
the geometric pattern of samplings is the same within each pixel, as a
result of the displacement pattern illustrated in Fig.\ 3.)  These
samplings approximate the continuous two-dimensional PSF.
Figure~\ref{datapts_slice} shows the value of each of these estimates
along two $x$-slices.

[FIG.\ 5 HERE]
[FIG.\ 6 HERE]

Our immediate task is to distill from this multitude of samplings the
values of the 21$\times$21 effective-PSF grid points which best
represent them.  (Let us ignore, for the moment, the fact that because
of the PSF's spatial variability each star actually samples a slightly
different PSF.  We will discuss in Section~\ref{varpsf} how to
generalize the following procedure to a spatially variable PSF.  In the
present context the only difference is that we take our samplings from
stars in only a part of the chip rather than from the whole chip.)

The process of determining the array of grid points from the samplings
is still another iterative one.  We first subtract from each sampling 
the most recent estimate of the ePSF at that point, 
$\psi_E(\Delta x,\Delta y)$, so that each data point is now a residual 
between the estimate and the current model.  In the first iteration, 
the PSF will be null, so that the residual will be the sampling itself.  
For each grid point we then take all the residuals within 0.25 pixel 
in $\Delta x$ and $\Delta y$ (shown by the heavy square in 
Figure~\ref{datapts}), find the average residual, iteratively reject 
those that are more than 2.5 sigmas away from the mean, and adjust the 
grid point by this amount.  The $\sigma$-based rejection procedure 
results in a robust estimate of the average residual for each grid 
point, so that we are insensitive to stars with undetected neighbors 
or to individual star images affected by cosmic rays.  

Once we have adjusted all the grid points, we then smooth them.  For
the interior points (solid dots) we use the following 5$\times$5
least-squares quartic kernel of smoothing coefficients:

\begin{equation}
\left[
\matrix{~~0.041632& -0.080816& -0.078368& -0.081816&~~0.041632 \cr
         -0.080816& -0.019592&~~0.200816& -0.019592& -0.080816 \cr
        ~~0.078368&~~0.200816&~~0.441632&~~0.200816&~~0.078368 \cr
         -0.080816& -0.019592&~~0.200816& -0.019592& -0.080816 \cr
        ~~0.041632& -0.080816& -0.078368& -0.081816&~~0.041632 \cr }
\right].
\label{kernel}
\end{equation}
(
Again, the number of points and the order of the fit are choices made
after trial and error, to satisfy the criteria of faithful smoothing
without creating artifacts.)  For the outer points (open squares), we
take the nearest 5$\times$5 grid points and perform a simple linear
least-squares fit of the log of the grid-point value against the
distance of the grid point from the center of the whole grid-point
pattern (the ``radius'').  We evaluate the linear function at the radius
of the grid point, and use the antilog of this as the value of the
grid point.

We smooth the outer points (open squares) 
with a linear kernel based on the closest 4$\times$4 grid points.  The 
coefficients in the three cases (corner, edge, or next to the edge) are 
easily derived.

We smooth the outer points (open squares) 
with a linear kernel based on the closest 4$\times$4 grid points.  The 
coefficients in the three cases (corner, edge, or next to the edge) are 
easily derived.

At the end of this procedure it is possible that the resulting ePSF
might not be centered properly.  [Recall that we require the central
pixel of the ePSF to have the symmetries 
$\psi(-0.5,\delta y) = \psi(+0.5, \delta y)$ and 
$\psi(\delta x,-0.5) = \psi(\delta x,+0.5)$, 
where $\delta x$ and $\delta y$ vary from $-$0.5 to 0.5.]  
To correct the centering, we simply offset the placements of all samplings 
so that the subsequent ePSF derivation will be properly centered.  We 
do this by the shift 
\begin{equation}
  \Delta x \rightarrow \Delta x + 
     { \psi_E(0.5,0.0) - \psi_E(-0.5,0.0)
       \over { 
                { \partial \psi_E \over { \partial x }} ( 0.5,0.0) + 
                { \partial \psi_E \over { \partial x }} (-0.5,0.0)
             } 
     }
\label{recensam}
\end{equation}
(and a corresponding expression for $\Delta y$), where the derivatives
are evaluated by simple differencing of straddling grid points.  This
shift is determined at the edges of the central pixel, but is applied to
every sampling point.

We iterate this adjust-and-smooth process several times, subtracting
each new model and smoothing the newly adjusted points.  The result is a
set of grid points that vary smoothly, are centered on the brightest
pixel, and are a good fit to the samplings.  With the present data set,
we found 5 iterations of this inner loop to be sufficient.

Finally, before using the ePSF in the next stage, we first re-scale it
so that a unit-flux star that is centered on a pixel has a total flux of
exactly 1.00 within its inner 5$\times$5 pixels (as explained in Sec.\
\ref{think_epsf}).

\subsubsection{Treating the spatial variability of the PSF}
\label{varpsf}
Thus far we have referred to ``the'' PSF, as if there were one universal
PSF which applies everywhere on the chip.  In most cameras (WFPC2
included), the PSF varies slightly, but significantly, with location on
the chip.  Mathematically, $\psi_E = \psi_E(\Delta x,\Delta y;x,y)$.  We
looked initially into allowing the PSF to vary quadratically with
position on the chip, analogously to the procedures in DAOPHOT or in
Guhathakurta et al.\ (1992); however we found that such a PSF did not
provide an adequate fit to the stars over the entire chip
simultaneously.
  
After some experimentation, we found that the following simple 
procedure gives the PSF the needed flexibility.  We determine nine ``fiducial''
PSFs as shown in Figure~\ref{fiducial}:\  one at the center, four at the 
corners, and four in the middle of the edges.  Recall that a ``PSF'' 
consists of an array of values for the $21 \times 21$ grid points.  To 
specify the PSF at any specific location in the image, we use simple
bi-linear interpolation among the four nearest PSFs:  the central one,
two at the edges, and one at a corner.  We investigated more complicated
interpolation schemes (radial functions with linear or quadratic dependences),
but found that simple linear interpolation provided equally good fits to 
the stars.  Each time we create a PSF for an individual star, this 
interpolation is done for each of the 441 grid points.

[FIG.\ 7 HERE]

We solve for each of the 9 fiducial PSFs as prescribed in
Section~\ref{sam2psf}, using for each PSF only the samplings within the
appropriate region (delineated by the dotted lines in Fig.\
\ref{fiducial}).  Since the PSF is not constant even within a region,
when subtracting the current model PSF from each sampling to form a
residual we subtract the properly interpolated PSF for the contributing
star's location, $\psi_E(\Delta x,\Delta y;x_*,y_*)$, using the other
fiducial PSFs for the interpolation, while adjusting only the one for
the region in question.  In this way we derive each fiducial PSF only
from the stars in its own region, while always comparing the samplings
from each star with our current best estimate of the ePSF that applies
to its location.
 
\subsection{Stage 2:  Fitting stars with an effective PSF}
\label{fitPSF}
Now that we have an improved estimate of the effective PSF, we can use it
to make improved fits to the individual star images.  The fitting of the 
effective PSF to each star image is a least-squares process.  The relation 
between the pixel values and the the position and flux is non-linear, 
however; and direct application of the non-linear least-squares algorithm 
requires first derivatives of the effective PSF.  Rather than treat the 
least-squares solution generally, which would necessitate a different matrix 
inversion for each star, for computational ease we have linearized the 
$\chi^2$-minimization equations and solved them analytically (see below).  
It is not of course necessary to adopt the following equations to fit a 
PSF to star images; but this method is very convenient and arrives
easily at the least-squares result, which is unique.

Essentially all the information about where a star is located is 
contained in the inner few pixels of the stellar image, where
the gradient of the PSF is highest.  We therefore confine our
fit to a fitting radius of $\sim$1.5 pixels.  In the interest of (1)
fitting only whole pixels, (2) having a roughly symmetric aperture with
respect to the star's center, and (3) having the aperture vary smoothly
with the presumed center, we adopt a weighting scheme similar to that
described and justified by Stetson (1987).  Pixels within 1.5 pixels 
of the star's center receive a full weight $w_{ij}$ of 1.0, while those 
beyond 2.0 receive zero weight.  The weight varies linearly from 
1.0 to 0.0 for distances between 1.5 and 2.0 pixels.  Note that this 
positional weight $w_{ij}$ is independent of Poisson considerations,
which we will include separately in our formulation.  Such a procedure
obviously involves some prior knowledge of the star's position, so that
the aperture itself must change slightly with each iteration.  The third
point above assures that it will not vary chaotically as we converge
upon $x_*$ and $y_*$.

With the fitting radius just described, we do not actually need to use values of
the ePSF from the entire 5$\times$5-pixel region over which we have
tabulated its value.  We solve for it over that entire region,
however, because the need to use derivatives of the ePSF brings in 
pixels farther from the center, and the proper smoothing of our ePSF 
grid points also requires values far out.

When we set out to measure a star, we first construct a PSF appropriate
for its location in the chip, by means of the prescription given in
Section~\ref{varpsf}.  The value of this effective PSF is tabulated at
the quarter-pixel grid points that we have described.  We then construct
a similar 2-D table for the $x$ and $y$ derivatives of the PSF (needed
below), by simple differencing of the smooth PSF grid.  This procedure
requires evaluating the PSF at intermediate locations within the grid,
for which we use a bi-cubic spline.  For the derivatives, we use simple
bilinear interpolation.

Once we have chosen the array of pixels to be fit, we minimize the following 
expression for $\chi^2$ with respect to the parameters $f_*$, $x_*$, and 
$y_*$:
\begin{equation}
  \chi^2 = \sum_{(i,j) \in {\rm AP} }
      {   w_{ij} \bigl[ (P_{ij}\!-\!s_*)-f_*\psi_E(i\!-\!x_*,j\!-\!y_*)\bigr]^2
%-----------------------------------------------------------
                    \over{  g P_{ij} } }\, ,
\label{chisqeq}
\end{equation}
where the sum is over all pixels $(i,j)$ within the fitting aperture, 
$P_{ij}$ is the value of the pixel at $(i,j)$, $s_*$ is the value of 
the sky, $\psi_E(\Delta x,\Delta y)$ is the value of the effective PSF 
at that offset, and $g$ is the gain ratio, typically $\sim$7 for WFPC2.  
The denominator normalizes the residual of each pixel for the expected 
Poisson noise in that pixel, so that what we have is a true chi-squared.  
In what follows, we write $q_{ij}$ for $1/gP_{ij}$.

The parameter $f_*$ appears linearly in this equation, while $x_*$ and $y_*$ 
enter non-linearly through $\psi_E$.  For a given $(x_*,y_*)$, we can easily 
solve for the flux $f_*$ that minimizes Eq.\ (\ref{chisqeq}):
\begin{equation}
       f_* = { \sum w_{ij} q_{ij} P_{ij}^{\prime} \psi_{ij} 
       \over { \sum w_{ij} q_{ij} \psi_{ij}^2 } }.
\label{fluxeqn}
\end{equation}
Here, we have defined $P_{ij}^{\prime}$ to be the sky-subtracted pixel 
value, $P_{ij}\!-\!s_*$, and $\psi_{ij}$ is shorthand for 
$\psi_E(i\!-\!x_*,j\!-\!y_*)$.  Once we have solved for the flux, we compute
the residuals $R_{ij} = P_{ij}^{\prime} - f_*\psi_{ij}$.  We then solve for 
the position parameters $x$ and $y$ via an iterative Newton--Raphson method, 
where $x_*^{[n\!+\!1]} = x_*^{[n]} + \delta x_*^{[n]}$, and
\begin{equation}
     \delta x_* = { \biggl[ \sum w_{ij} q_{ij} R_{ij} \!
                         \left({\partial\psi_{ij}\over{\partial x } }\right) 
                  \biggr]  \!
                  \biggl[ \sum w_{ij} q_{ij} \!
                         \left({\partial\psi_{ij}\over{\partial x } }\right)^2 
                  \biggr]
                - \biggl[ \sum w_{ij} q_{ij} R_{ij} \!
                         \left({\partial\psi_{ij}\over{\partial y } }\right) 
                  \biggr] \!
                  \biggl[ \sum w_{ij} q_{ij} \!
                         \left({\partial\psi_{ij}\over{\partial x } }\right) \!
                         \left({\partial\psi_{ij}\over{\partial y } }\right) 
                  \biggr]
             \over{
           f_* \biggl(
                  \biggl[
                     \sum w_{ij} q_{ij} \!
                     \left({ \partial\psi_{ij}\over{\partial x}} 
                     \right)^2
                  \biggr] \!
                  \biggl[
                     \sum w_{ij} q_{ij} \!
                     \left( {\partial\psi_{ij}\over{\partial y}} 
                     \right)^2
                  \biggr]
                - \biggl[
                     \sum w_{ij} q_{ij} \!
                     \left( {\partial\psi_{ij}\over{\partial x}} 
                     \right) \!
                     \left( {\partial\psi_{ij}\over{\partial y}} 
                     \right)
                  \biggr]^2
             \biggr)
                  } }.
\label{xadjeqn}
\end{equation}
The equation for $\delta y_*$ is analogous.  We alternate between solving 
for $f_*$ and for $(x_*,y_*)$.  Convergence is typically reached in very 
few iterations.  With the PSF and its derivatives already tabulated ahead of 
time, the center-finding procedure is very fast.  Typically, we can measure 
2500 stars per minute on a Sparc Ultra 5.

\subsection{Stage 3:  Position analysis and refinement} 
\label{posref}
The result of the second stage is an improved list of the raw positions 
of each star as measured in each image.  We use our transformations 
to transfer all of these raw measurements into a common reference frame
so that we can combine them and inspect the residuals.  (It is of course
crucial to have multiple ditherings to give residuals that can be used 
to test for pixel-phase error.)  If the measurements are free from bias, 
then a plot of the residuals against pixel-phase (as in Fig.~\ref{ppe}) 
will show only random errors.  However, if any significant pixel-phase 
error remains, we will need to iterate again.

The presence of pixel-phase error indicates systematic errors in the raw
positions we just measured.  If we were to take these biased positions
and use them to create new PSF samplings and solve once again for the
PSF, then we would arrive a PSF that is as biased as before.  Thus in
order to remove the bias we must introduce some new information.  This
new information comes from the multiple observations we have of the same
star at different dither positions.  The same averages that we use above
to examine pixel-phase error can be used to correct it, not in a purely 
empirical way, but rather by going to the root of the problem and 
{\it removing} it, by the accurate PSF determination that we are 
describing here.

To derive an improved placement for each PSF sampling, we transform the
mean position of each star from the reference frame into each of the
individual frames where it has been imaged.  We then have a more
accurate (i.e., less biased) estimate of which point in the effective PSF
each of the star's pixels samples.  When we then solve again for the
ePSF it will be more accurate, and can be used to measure positions that
are in turn more accurate and less biased.  

In a similar way, we average the fluxes from the multiple observations
to arrive at an improved value for the flux of each star in each image
(allowing of course for any differences in exposure time).  Quite in
parallel with the astrometry, this enables us to allow for the variation
of total detected flux with pixel phase (Lauer 1999b), which can be $\pm
1.5$\%.  This variation is handled naturally and properly with the
effective-PSF formalism.

\subsection{Summary of the major points} 
\label {ePSF_summary}

The flow chart in Figure~\ref{flowchart} shows our overall procedure for
arriving at an accurate PSF.  Stripping away the details, we summarize here 
the major points of this section.

[FIG.\ 8 HERE] 

(A) We recognize the role of the effective PSF, which will have to be
    evaluated at arbitrary points but will never have to be integrated.
    We choose to tabulate this continuous function at a well-sampled 
    mesh of points.

(B) We derive the ePSF by using every pixel value of every well-exposed
    star; each pixel value contributes information at a particular
    point in the ePSF.

(C) We allow for the spatial variation of the PSF by solving for 9 fiducial
    PSFs placed at extreme points in the image, so that we can interpolate 
    among them to get the locally appropriate PSF at any place on the chip. 

(D) By iterating between PSF determination and measurement of star
    positions, we improve successively the placement of the individual
    pixel values in the building up of the PSF, and the fit of the PSF
    to the star images.

(E) The crucial part of this procedure is the transformation of the mean
    positions back into the individual images, which allows each image
    to profit from the averaging of all of them.  This linch-pin step removes
    the fundamental degeneracy caused by detector undersampling.

In this way we derive from the complete set of pixel values in the stellar
images the nine sets of 441 grid points that represent the effective PSF 
throughout the image. 

\subsection{Result of iterations}
\label {itern_results}

Figure~\ref{psfvsit} shows graphically the progression of ePSF samplings
and pixel-phase errors with iteration, for stars in the central region of
the WF2 chip.  In the first column the samplings come from centroid positions 
and 3$\times$3-pixel-aperture fluxes.  These raw samplings are not well fit 
by a smooth model.  The pixel-phase errors at the bottom show the bias in 
positions measured with this model.  To illustrate the importance of
averaging, in iteration $2^{*}$ we take these positions without
averaging, re-sample the PSF, and derive a smooth model that does seem
to fit the samplings reasonably well.  However, we find that the
positions measured from this PSF are as biased as before.  The smooth
fit of the model to the samplings does not guarantee a correct
(bias-free) model.  The pixel-phase bias is clearly visible even in the
top panel, which shows the overall distribution of samplings.

[FIG.\ 9 HERE]

The rightmost two columns show the standard progression of our
iteration.  For the column labeled ``2'' we take the positions after the
first iteration, average them with the positions for the same stars from
the other dither-pointings, then use these {\it average} positions to
re-sample the PSF.  The locations of these new samplings are shown at
the top; the removal of the bias is remarkable.  These new samplings
suggest a PSF which is shaped quite differently from that in iteration
$2^{*}$.  When we fit these samplings with a smooth model, we find
model-residuals that seem a little inferior to those for iteration
$2^{*}$.  However, when we use this new model to fit stars, we find that 
the pixel-phase error goes down by more than a factor of two.  As we
iterate yet again, the samplings appear even more evenly distributed,
and the resulting pixel-phase errors are yet again diminished.  We 
find that after 5 or so such iterations the bias is typically
reduced to a negligible level.

[FIG.\ 10 HERE]

Figure~\ref{ppe_vs_it} shows the pixel-phase-error progression
for all 9 regions of the WF2 chip.  Plots for the other WF and PC chips 
are very similar.  This figure clearly demonstrates that our spatially 
variable effective-PSF model coupled with our iterative procedure have 
reduced the pixel-phase bias to a negligible level ($<$0.002 pixel) for 
{\it all} regions of the chip.  To put this in perspective, we note that 
the random error in measurement for a typical star is $\sim$0.02 pixel 
in each coordinate.  Because of our freedom from systematic error, the  
average positions from 15 images will be better by a factor of $\sqrt{15}$ 
and have a typical error of only 0.005 pixel.  We believe that an $N$ 
considerably larger than 15 will lead to a correspondingly improved 
accuracy, since, as we will show in Section \ref{indiv_comp} and 
Fig.\ \ref{ppe_vs_it}, we are able to remove pixel-phase error to a 
level considerably below 0.005 pixel.

\subsection{The final PSFs} 
\label{final_psfs}

Figure \ref{ppefix} contrasts strikingly with Fig.\ \ref{ppe}, and shows
how well pixel-phase error has been removed from our measurements.  The
small ``systematic error'' that remains is at the level that would be
expected from statistical fluctuations due to the limited sample size.

[FIG.\ 11 HERE]

In the left half of Figure~\ref{show_psfs} we show contour plots of our
final PSFs for the PC (shown for the $3\!\times\!3$ set of regions of
the image), and similarly, below, for the WF2 chip.  In the right half
of the figure we show cuts along the $x$ and $y$ axes of the PSFs in the
central row and column of the set at the left.  It is clear that the
variation of the PSF has a strong radial dependence:\ as we go from the
center of a chip outwards, the center of the halo moves with respect to
that of the core.  We note that all these PSFs have been rigidly
centered so that $\psi_E(-0.5,0.0) \equiv \psi_E(0.5,0.0)$ and
$\psi_E(0.0,-0.5) \equiv \psi_E(0.0,0.5)$, which also facilitates
interpolation among models.  The smoothness of these contour plots
demonstrates the effectiveness of our smoothing kernel.

[FIG.\ 12 HERE]

\subsection{Comparison of individual images}
\label{indiv_comp}
We have derived our PSF by combining information from 15 images.  The
question arises, however, {\it do} all the images have exactly the same PSF?
In principle, the answer must be negative.  In actual fact, indeed, we
found that the image-to-image transformations generally involved small
differences of scale.  The time sequence of these showed that they were
clearly due to the phenomenon called ``breathing'', in which the \hst\
tube length varies around each orbit as a result of changing insolation.
This must lead to small changes of focus, which will undoubtedly affect 
the PSF.

If the PSF in one of the images differs from the average PSF but we use
the average PSF to measure the stars in that image, we would expect these
measurements to contain pixel-phase errors caused by the PSF mismatch.
Figure~\ref{probexy} shows the net pixel-phase errors for each of the
15 images of the 1st-epoch sample.  All of these errors are well below 
0.005 pixel.  Thus even for our demanding purposes it is safe to use
the average PSF when measuring each image.  

[FIG.\ 13 HERE]

The bottom-right plot in Figure~\ref{probexy} shows that we tend to
measure stars slightly better ($\sim$15\%) when a star is centered near
the boundary between two pixels than when it is centered in a pixel.
Intuitively we would expect this to be the case, since when a core is 
centered on a pixel boundary any slight change in position of the star 
results in a maximal redistribution of light from one pixel to another.  
Still, this is a small effect and we find that we can measure stars 
well wherever they fall upon the pixel grid.

Since it is safe to treat the PSF as constant within an epoch (i.e.,
over several orbits), it is appropriate to ask about the long-term
stability of the PSF.  We reduced the 2nd-epoch (1997) images with the
1st-epoch (1995) PSF and examined the measurements for pixel-phase
error.  Figure~\ref{cross_psf} shows that the PSFs must be appreciably
different from one another, both in overall sharpness and in
spatial-variability characteristics.  It is therefore crucial to 
obtain a separate PSF from each epoch observed and to verify any 
assumptions of PSF constancy.  It is further clear that ``library'' 
PSFs will not be adequate for high-precision astrometry.

[FIG.\ 14 HERE]

\clearpage

\section{Effective dithering strategies for astrometry}\label{dith}

\subsection{Planning observations}\label{optdith}
The dithering needs of astrometry are quite different from those of
image reconstruction.  In the latter case, we have an unknown and 
possibly complicated scene which we would like to sample as 
well as possible.  Simple interlacing (as described in Lauer 1999a) is 
the most straightforward and complete way to create a well-sampled map 
of the effective scene.  The situation is quite different for astrometry, 
where every star is merely a replica of the same PSF (spatial variations 
of the PSF aside).  

Our objective in dithering is to permit an accurate reconstruction
of the PSF.  Because each star is a differently centered representation 
of the same PSF, we will typically have hundreds or thousands of different 
realizations of the PSF.  There is no question that the PSF will be 
``well-sampled'' and over-constrained.  However, we still need dithering 
to tell us how each of the stellar images in each frame samples the 
PSF, so that we can construct a PSF that is free from bias.  For this 
purpose, there is no ``ideal'' dither pattern.  

We have not yet done much experimentation on how small a number of
dither pointings will suffice, but it is likely that a 2$\times$2
pattern with 0.5-pixel steps would be adequate.  Because stars fall at
random phases, we would have in the equivalent of Fig.\ \ref{ppe} a
large number of pairs of residuals half a cycle apart, {\it at all
phases\/}, and this is enough to construct the curves of pixel-phase
error.  Here again, as in Section \ref{relation}, we emphasize that our
problem is not removal of aliasing; what matters is rather the denseness
of the samplings illustrated in Fig.\ \ref{datapts}.  It is this last
consideration that argues for using as many different dither pointings
as is reasonably possible.  Repeating an exposure at the same dither
pointing would merely produce additional samplings at the same points in
the ePSF as before, and would not improve the derivation of the ePSF.
On the other hand, it is the science program that should dictate the
number of exposures---as long as there are at least 2$\times$2 dithers.
After this number is chosen, then one should specify a dither pattern
that is spread out relatively evenly over the 2-D pixel-phase space, but
there is no need to create a perfectly regular pattern.

%% New paragraph added:
Thus the minimal needs of astrometry, a 2$\times$2 dither, can be less
than those of image reconstruction, where Lauer (1999a) states that for
the F555W ($V$) filter he needed 3$\times$3 in the WF chips (and
presumably even more for the sharper PSFs that are formed in the UV).
As we have argued, however, when the science program allows a larger
number of exposures it is also desirable to use a large number of
dithers.

An additional requirement is to achieve a good dither in the WF and 
the PC simultaneously, given their different and incommensurable pixel
sizes.  We solve this problem by choosing a phase pattern that satisfies
our criterion as well as possible for the WF, in which the more severe
undersampling makes dithering more critical, and adding whole-WF-pixel
offsets to the dithers so as to make those in the PC as close as
possible to ideal.  We have no set procedure for this, but choose the
whole-pixel offsets by trial and error until they result in a good
pattern for the PC.

[FIG.\ 15 HERE] 

Figure~\ref{dither15} shows an example of a 15-way dither, the
within-pixel phases at the top and the total dither amounts at the
bottom.  This particular example was chosen to fit a constraint of \hst\
program construction for this target:\ there were to be 5 exposures in
each of 3 orbits.  The exposures are arranged to fit into WFPC2's
procedure of line-dithering, which is constrained to consist of equally
spaced exposures along a straight line.  For a different number of
exposures, or different restraints on their grouping, good dither
patterns are not hard to construct. 

One might wonder whether the considerable geometric distortion in WFPC2
might defeat any attempt to dither evenly.  The answer is that because
each star is shifted by a small number of pixels within the dither
pattern, the {\it relative} offsets of the 15 images of each star are
essentially the same everywhere on the chip.

\clearpage

\subsection{Dealing with non-optimal dithering}\label{nonoptdith}

We note that while dithering clearly provides the best and most
straightforward way to derive an accurate model of the PSF, it is by no
means the only way to do so.  In the absence of a dithered set, there
are still some tricks that can be played to tease apart the position-PSF
degeneracy, whose removal, we have emphasized, is the major challenge 
to the astrometry of undersampled images.

When it comes to measuring proper motions, we often have the freedom to
plan the second-epoch observations, but must deal with a first epoch
whose observations have already been taken.  To plan and reduce the
second epoch, we recommend the prescription given in the preceding
section.  But special care must be taken to reduce the first epoch, if
it is not well dithered.

We note that sometimes even though no deliberate dither exists for the
first epoch, there may yet exist images of the same field taken through
different filters.  The different filters often have different plate
scales, so that there may be an effective dither between the two images.
This can be exploited by solving for different PSFs for the two images,
while constraining the stars to have the same relative positions (since
the images were taken at the same epoch).  The iterative scheme
described in Section~\ref{find_psf} could easily be modified for this
situation.

If the first epoch is completely undithered, there is still a possible
remedy.  We can make use of the fact that nature should distribute stars
randomly with respect to the pixel boundaries, so that if in
constructing the PSF we find our pixel phases to be biased (as in the
upper left panel of Fig.~\ref{psfvsit}) we could introduce a general
correction to flatten out this distribution.  We have experimented with
such adjustments and find that we can reduce the systematic pixel-phase
error to less than 0.01 pixel (from 0.03 pixel or so).  Such a
correction is naturally dependent upon having a reasonably large sample
of stars, so that the statistical assumptions of a flat pixel-phase
distribution will be valid.  It is also important to exclude {\it a
priori} non-stellar objects (galaxies, or peaks created by cosmic rays
or warm pixels) from the PSF lists, since any bias in the PSF-measured
positions for them would be irrelevant to a construction of the PSF.

Some amount of iteration may be necessary in both of the above schemes.
We can test the accuracy of the PSF model by searching for trends with 
first-epoch pixel phase, in the final inter-epoch displacements.  Such a 
test should at the very least allow us to put an upper limit on any 
remaining systematic errors.

\clearpage

\section{Extensions of the effective formalism}
\label{ePSFextend}
We have introduced three important ideas in this paper:\ the value of
using the effective PSF, the incorporation of pixel values from {\it
all} well-exposed star images to derive a smooth, parameter-free model
of the PSF, and iteration between PSF shape and star positions.  All of
these ideas can be extended beyond the present application of WFPC2
astrometry.

Whenever one must deal with intrinsically pixellated data, an effective 
approach will provide for economy of computation and transparency of 
analysis.   The ePSF approach is especially useful in the case of 
undersampled data, where it allows one to model directly the way flux
is transferred from one pixel to another as the ``center'' of a point source
moves with respect to pixel boundaries.  Since many undersampled wide-field 
detectors exist in ground-based astronomy, these ePSF techniques 
developed for \hst\ are equally relevant on the ground.  Use of the 
effective PSF itself is, of course, no less appropriate for well-sampled data.

Our parameter-free, grid-based modeling approach is flexible enough to
deal with vastly different PSFs.  It frequently happens that because of
guiding problems, the PSF is elongated in some direction, which is not
necessarily along the rows or columns.  Standard, analytical-model-based
photometry routines have difficulties measuring such images.  Our model
is flexible enough to deal naturally with such ``ugly'' PSFs.  We make
no assumptions about the shape of the PSF, except that it is relatively
smooth (which we can accomplish by choosing an appropriate smoothing
kernel).  Aside from the obvious slight degradation of resolution along
the elongation direction, we find that we are able to analyze these
images as easily as perfectly guided images.

In fact, our flexible ePSF approach would be ideal for analyzing images
that are intentionally trailed.  Such images could in principle offer
ultra-precise positioning information in a direction orthogonal to the
trailing direction.  Without trailing, we believe that there remains
some fundamental limit to our ability to determine the ePSF, and
therefore a limit of $\sim$0.002 pixel to our systematic accuracy in an
individual measurement (see Fig.\ \ref{ppe_vs_it}).  Trailing might
reduce this remaining error significantly.

A final recommendation for the effective approach is that it is the
ideal PSF to use for image reconstruction.  Lauer (1999a) makes the
point that there is no reason to treat the integration over pixels and
the instrumental-PSF convolution separately---mathematically the image
is literally the convolution of the fully resolved scene with the effective
PSF, sampled at the pixel centers.  Thus, if one desires to reconstruct
the scene accurately, the best approach is to deconvolve the observation 
with the effective PSF.  Having an accurate effective PSF is therefore 
the first step.

\clearpage

\section{Summary}
\label{summary}
The key to accurate astrometry, especially in undersampled images, 
is to derive a point-spread function of the highest accuracy possible.
We demonstrate how a less-than-adequate PSF leads to pixel-phase error, 
i.e., systematic position errors that depend on the location of a star 
with respect to pixel boundaries; and we show that an accurate PSF 
eliminates pixel-phase error. 

Consideration of the practical use of a PSF leads to the concept of 
the effective PSF, a continuous function that is the convolution of the 
PSF that the telescope delivers to the focal plane with the spatial 
sensitivity function of an individual pixel.  This ePSF has the 
practical virtue that the expected value of any pixel in any star image 
is the product of the value of the ePSF at the location of the center of 
that pixel with respect to the center of the star image times a 
factor that expresses the brightness of that star.  Once the ePSF is 
tabulated, no integration over a pixel area ever needs to be performed.

The ePSF can be evaluated accurately by a suitable combining of the
observed pixel values of all stars bright enough to be used in this
process, which we have described in some detail.  The evaluation process
depends on an iteration in which the shape of the ePSF and the measured
positions of the individual stars in a set of dithered images are
alternately improved, in a succession that converges to accurate values
of both data sets.  What is essential to the convergence of the
iteration is the ability to average the positions of each star at
various dither offsets.

We describe a method of interpolation between ePSFs derived for 9 
separate parts of a field, by which an accurate ePSF can be found for a 
star anywhere in the field.

After a brief discussion of strategies that will achieve good dither
patterns, we describe a modification of our methods that allows fairly
complete removal of pixel-phase error in material that lacks dithering
information, and we discuss ways in which our approach can also benefit
the treatment of images that are well sampled.

We must caution, however, that our methods rely on having a reasonably
large number of stars in each field and are therefore not applicable to
sparse star fields.  In order to reconstruct a good PSF we should have
at least 100 well-measured stars in each chip.  For the purposes of
accurate differential astrometry, however, the number of stars needed
will depend greatly on the desired precision and on the quality of
distortion correction that is available.  We will take up these issues
in a subsequent paper, in which we will discuss the transformation of
coordinates from one image to another.

\acknowledgments 
We are grateful to the referee, Tod Lauer, for a number of suggestions
that considerably improved the quality of this paper.  We thank Adrienne
Cool for the DAOPHOT measurements that we used here, and we are also
grateful to William van Altena for discussions.  This work was supported
by grant AR-7993 from the Space Telescope Science Institute.

\newpage

\noindent{\bf Figure captions}
\bigskip
%\font\sc=cmcsc10

\figcaption{The histogram shows the pixel values for the innermost three
          pixels of an undersampled 1-D star profile.  The solid curve
          is a pure Gaussian model which, when integrated over the
          pixels, fits their values exactly.  The dotted curve is a
          composite of a sharper Gaussian with a small contribution from
          a broader one; integration over it also fits the pixels
          exactly.  The arrows show the locations of the peaks of these
          two PSFs, which are offset by 0.07 pixel.\label{simple_ex}} %1

\figcaption{Pixel-phase error in the measurements of 15 dithered but
         otherwise identical images in chip WF2, illustrated by plotting 
         residuals of $x$ position in a star image against the $x$ pixel phase 
         at which the star was centered in that image.  The graphs are for 9 
         regions of the image.  The quantities given after SYS and RAND refer 
         to the amplitude of the systematic trend (dark line) and the
         r.m.s.\ dispersion about that trend, respectively.\label{ppe}} %2

\figcaption{This plot shows graphically the relationship between the iPSF
         and the ePSF.  At any given point (e.g., the cross), the value of the 
         ePSF is equal to the integral over a pixel centered at the 
         corresponding point in the iPSF.  The smallest contour interval is
         0.1 dex.  The plot on the right shows three slices of the two 
         PSFs (iPSF solid, ePSF dashed).  The smoothness of both is clear.
         \label{ins2eff}} %3

\figcaption{At the left are the pixels in the vicinity of a star in one of
our images.  Crosses are pixel centers, and the circle is the center of
the star.  The right-hand diagram shows how each of these pixels samples
the effective PSF at a different place (marked by the crosses).  
\label{coords}} %4

\figcaption{ The regularly spaced symbols are grid points at which the
          effective PSF is to be tabulated; the small points are the
          locations of point-samplings of the effective PSF, as
          described in the text.  The heavy small square shows the
          region over which samples are examined to adjust the circled
          grid point.  The dashed box shows the region over which the
          indicated grid point is smoothed after all the grid points
          have been evaluated.\label{datapts}} %5
 
\figcaption{ This shows the correspondence between the many raw
          samplings of the effective PSF and the final iterated and
          smoothed grid values.  The solid line shows a spline through
          the grid points.  \label{datapts_slice}} %6

\figcaption{This figure shows the locations of the nine fiducial ePSFs and
the process of interpolation to find a PSF at a particular point on the
chip.  The regions $x<50$ and $y<50$ of each chip are shadowed by the
edges of the reflecting pyramid.  The dotted lines denote the region of
the image used to solve for each PSF.  \label{fiducial}} %7

\figcaption{A flow chart showing our iterative PSF-finding procedure.
         The inputs and outputs of each major operation are indicated
         along the connecting lines. 
         \label{flowchart}} %8

\figcaption{The progression of ePSF samplings and pixel-phase errors
with iteration.  The top row shows the locations in the ePSF that are
sampled by the stars.  The second row shows the raw ePSF samplings for a
slice along $\Delta y \sim 0$.  The third row shows the fractional
residuals of the above samplings from the best-fit smooth model.  The
bottom row shows the resultant pixel-phase errors after stars have been
re-measured with the smooth model PSF.  See text for a description of
the iterations.\label{psfvsit}} %9

\figcaption{ For the 9 regions of a WF2 chip, we show the average $x$ residuals
          binned in pixel phase for iterations 1 (open circles),
          2 (filled squares), 3 (open squares), 4 (crosses), and 12 (solid
          circles).  For most of the regions the systematic residuals become 
          negligible (0.002 pixel) after a few iterations.   After 12 
          iterations, the residuals in all regions are negligible.
          \label{ppe_vs_it}} %10

\figcaption{As is Figure \ref{ppe}, residuals in $x$ position are
plotted against $x$ phase for the 9 sections of the image, to
demonstrate how well pixel-phase error has been removed by deriving an
accurate PSF.\label{ppefix}} %11

\figcaption{ Contour plots of the effective PSFs obtained for the 9
          regions of the image, for PC1 (top) and WF2 (bottom) chips.
          We show here only the restricted region between $-1.0$ and 1.0
          pixels in $x$ and $y$ (the inner $2\!\times\!2$ pixels of the
          PSF), as this is where most of the information relevant to
          centering is contained.  The smallest contour interval is 0.1
          dex.  (We solved for the effective PSF, however, over
          $5\times5$ pixels.)  The plots to the right show slices
          through the $x$ and $y$ axes for the three PSFs in the central
          row and the central column, respectively.  \label{show_psfs}} %12

\figcaption{ The overall pixel-phase residuals for exposures 1 through
          15 of the first-epoch WF2 chip.  The typical residuals are
          $\sim 0.002$ pixel.  In the bottom right plot, we show the
          overall trend of r.m.s.\ residual with pixel phase.  Filled
          circles refer to the $x$ coordinate, open circles to
          $y$.\label{probexy}} %13

\figcaption{ The pixel-phase residuals that result from using the first-epoch 
          PSF to reduce the second-epoch data.  The sense of the variation 
          seen in the middle plot implies that the first-epoch PSF is 
          {\it sharper} than the second-epoch PSF.  \label{cross_psf}} %14

\figcaption{A good 15-point dither pattern.  The left side represents 
         the WF and the right side the PC.  At the bottom are the 
         actual offsets of exposures 1--15, in pixels; at the top 
         the whole numbers in the offsets have been removed, so 
         that only the pixel phases (i.e., the actual dithers) are
         depicted.\label{dither15}} %15 

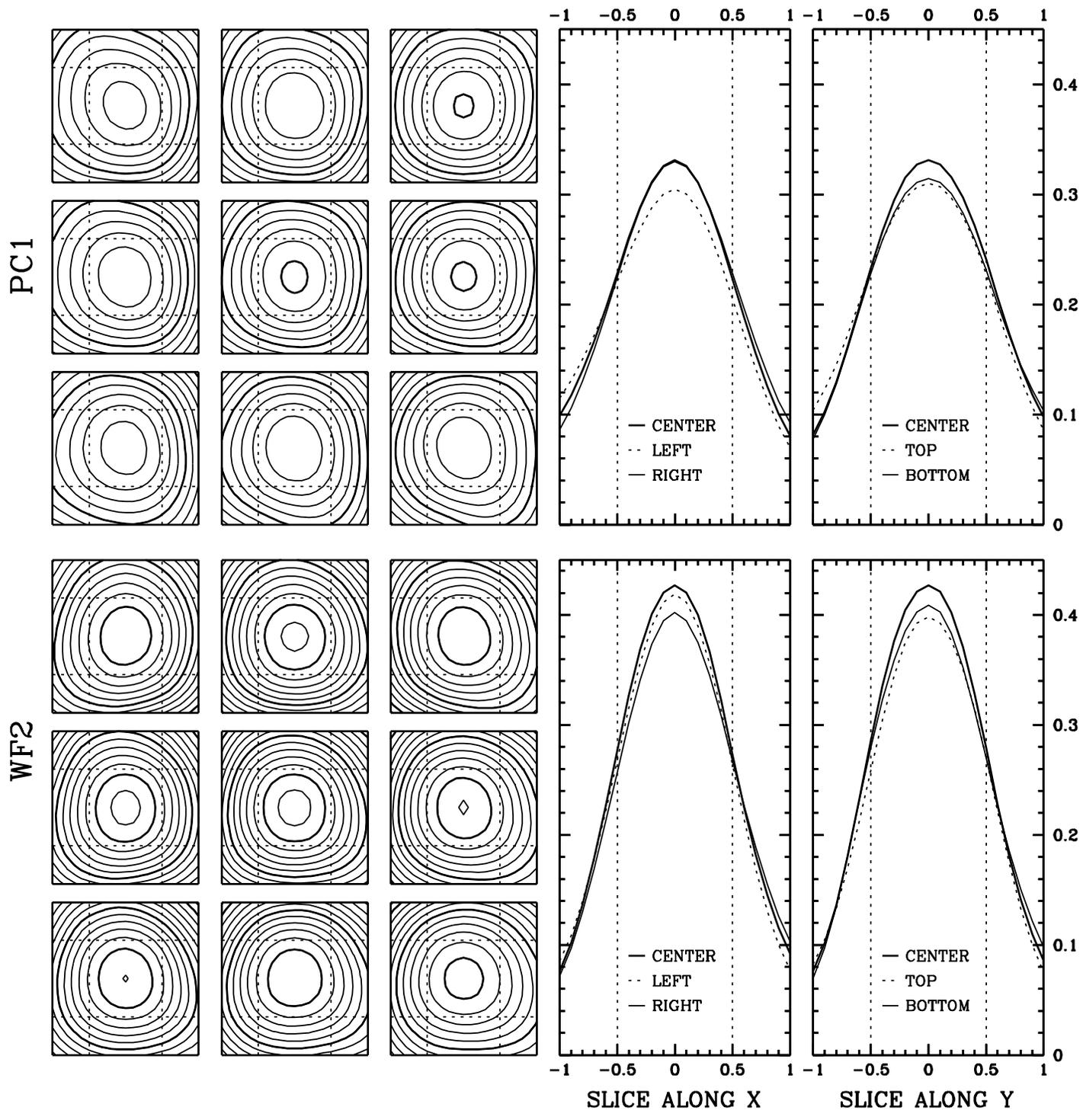
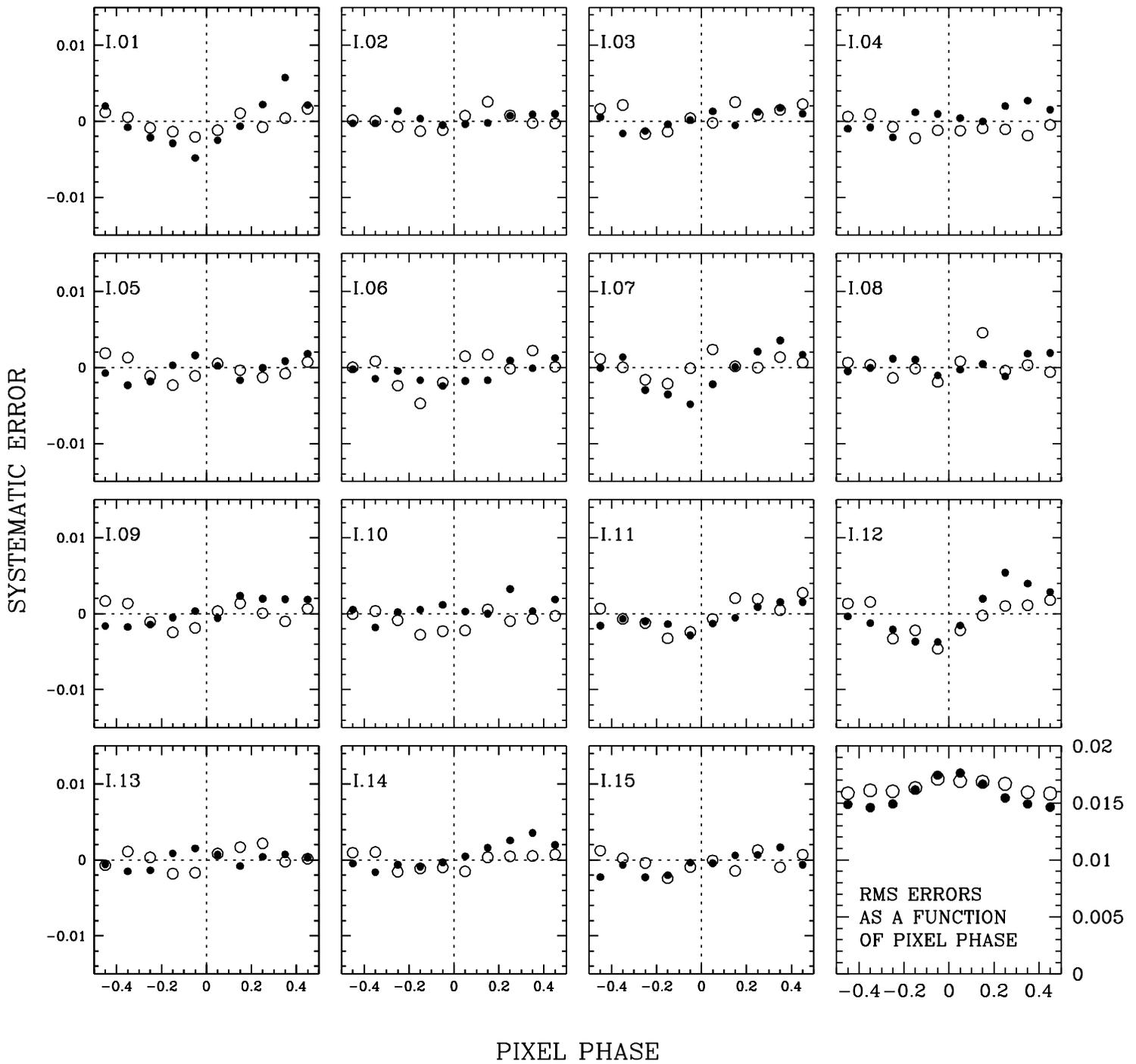
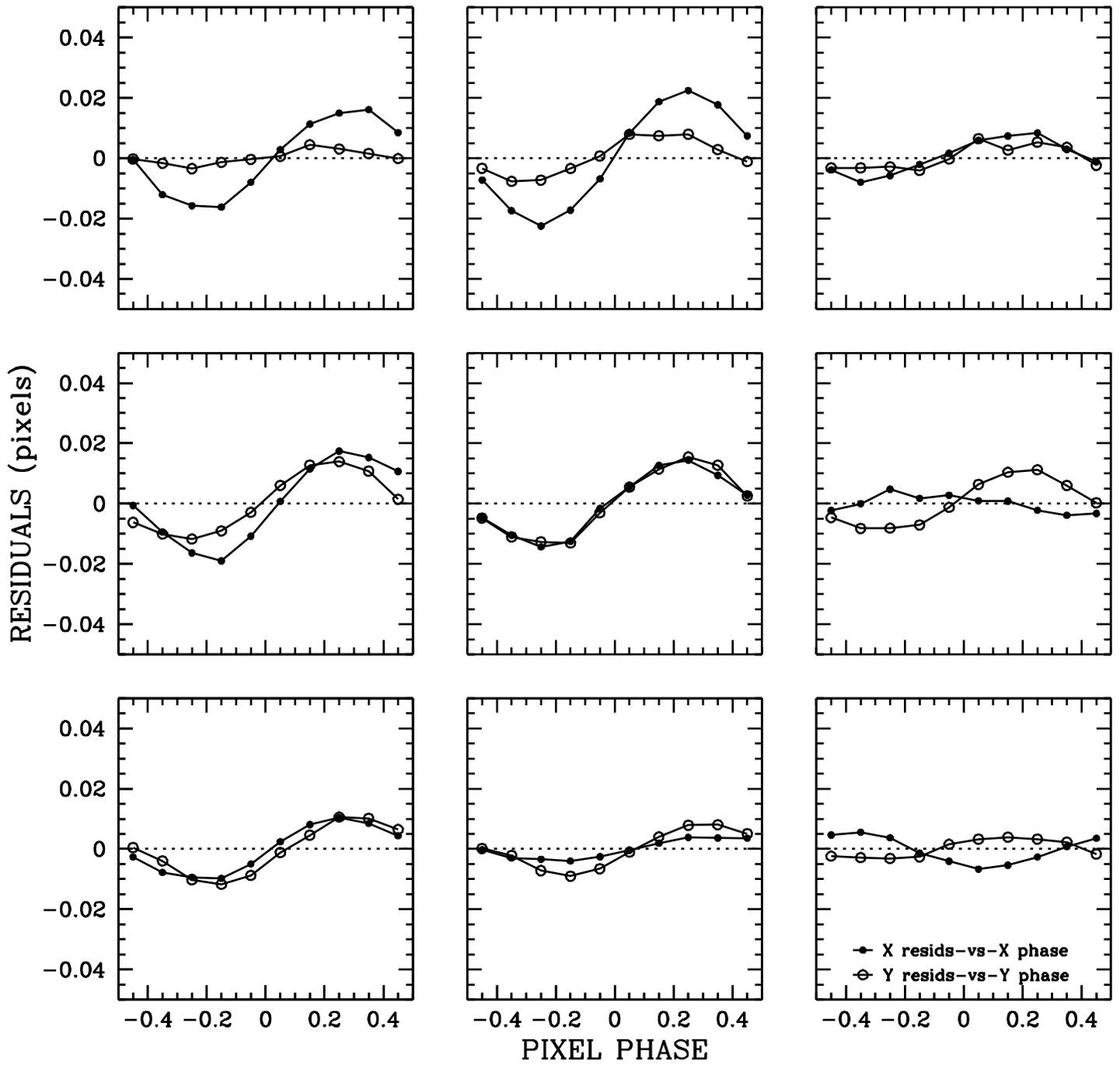
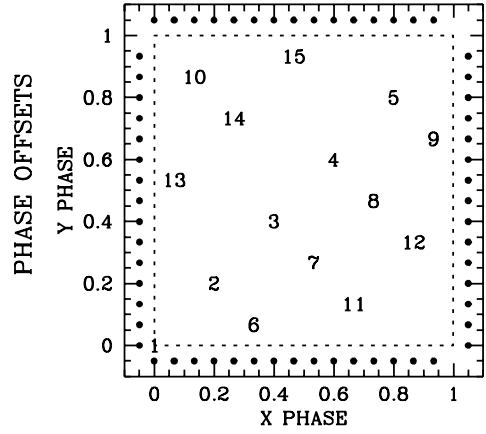
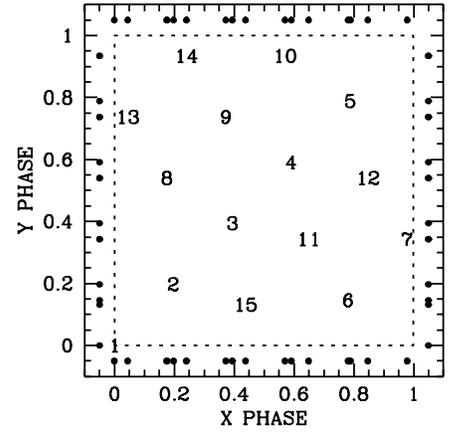
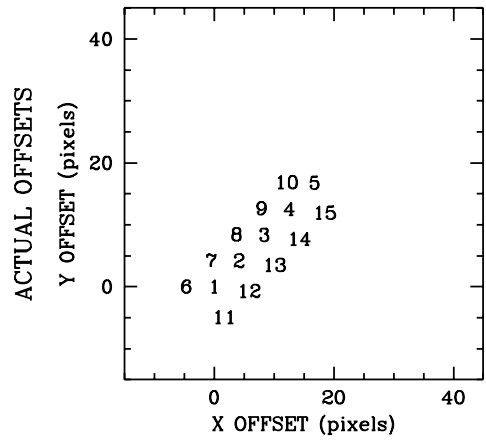
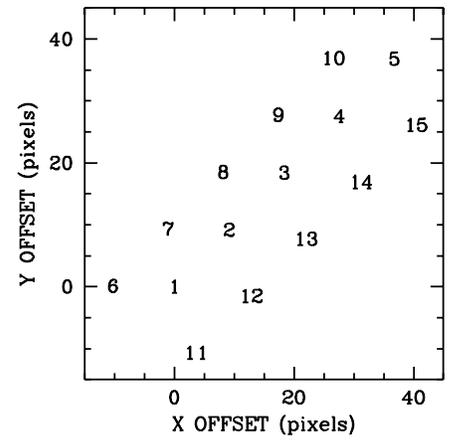
\end{document}